\documentclass[final]{aipproc}
\layoutstyle{8x11double}
\usepackage{amsfonts,amssymb,amsmath}
\newcommand{\w}{\omega}
\begin{document}
\title{The ghost propagator in Coulomb gauge}
\author{P.~Watson}{address=
{Institut f\"ur Theoretische Physik, Universit\"at T\"ubingen, 
Auf der Morgenstelle 14, D-72076 T\"ubingen, Deutschland}}
\author{H.~Reinhardt}{address=
{Institut f\"ur Theoretische Physik, Universit\"at T\"ubingen, 
Auf der Morgenstelle 14, D-72076 T\"ubingen, Deutschland}}
\begin{abstract}
We present results for a numerical study of the ghost propagator in Coulomb gauge whereby lattice results for the spatial gluon propagator are used as input to solving the ghost Dyson-Schwinger equation.  We show that in order to solve completely, the ghost equation must be supplemented by a boundary condition (the value of the inverse ghost propagator dressing function at zero momentum) which determines if the solution is critical (zero value for the boundary condition) or subcritical (finite value).  The various solutions exhibit a characteristic behavior where all curves follow the same (critical) solution when going from high to low momenta until `forced' to freeze out in the infrared to the value of the boundary condition.  The boundary condition can be interpreted in terms of the Gribov gauge-fixing ambiguity; we also demonstrate that this is not connected to the renormalization.  Further, the connection to the temporal gluon propagator and the infrared slavery picture of confinement is discussed.
\end{abstract}
\maketitle

The ghost propagator and its Dyson-Schwinger equation [DSE] are of central interest to studies of nonperturbative Yang-Mills theory.  In Coulomb gauge, there is known to be two types of infrared behavior: critical (with an infrared divergent dressing function) and subcritical (infrared finite dressing) \cite{Epple:2007ut}, similar to the situation in Landau gauge \cite{Fischer:2008uz}.  We demonstrate here how the existence of these different solutions is intrinsic to the standard ghost DSE in Coulomb gauge and how a particular pattern emerges.  This talk is based on the work of Ref.~\cite{Watson:2010cn}.

To solve the ghost DSE, we require two pieces of input: an Ansatz for the ghost-gluon vertex and the spatial gluon propagator.  In Coulomb gauge, the ghost-gluon vertex renormalization coefficient is nonperturbatively finite and since the vertex reduces to its tree-level form as one ghost leg momentum vanishes \cite{Watson:2006yq}, we can take $\tilde{Z}_A=1$.  Our vertex Ansatz is then to simply take the tree-level result (see Ref.~\cite{Watson:2010cn} for details of our conventions) as is standard:
\begin{equation}
\Gamma_{\overline{c}cAi}^{abc}(k_1,k_2,k_3)=-\imath gf^{abc}k_{1i}.
\label{eq:vert}
\end{equation}
The spatial gluon propagator is decomposed as follows:
\begin{equation}
W_{AAij}^{ab}(k)=\delta^{ab}\frac{\imath}{k^2}t_{ij}(\vec{k})D_{AA}\left(k_0^2,\vec{k}^2\right)
\end{equation}
where $t_{ij}$ is the transverse spatial projector and the dressing function reflects the noncovariance of the gauge.  As it turns out, the important quantity is not $D_{AA}$ itself, but rather the dressing for the equaltime propagator
\begin{equation}
D_{AA}^T\left(\vec{k}^2\right)=\imath\int_{-\infty}^{\infty}\frac{dk_0}{2\pi}\frac{D_{AA}\left(k_0^2,\vec{k}^2\right)}{\left(k_0^2-\vec{k}^2+\imath0_+\right)}.
\label{eq:dat}
\end{equation}
For definiteness we use the lattice results of \cite{Burgio:2008jr}, compatible with a Gribov formula \cite{Gribov:1977wm}, as input:
\begin{equation}
D_{AA}^T\left(\vec{k}^2\right)=\frac{1}{2}\frac{\sqrt{\vec{k}^2}}{\sqrt{\vec{k}^4+m^4}}.
\label{eq:gribov}
\end{equation}
Many other studies have reported similar results (see Ref.~\cite{Watson:2010cn} and references therein for a discussion), motivating the usage here.  The above form is already renormalized and it was found that the Gribov scale $m$ is independent of the renormalization scale $\mu$.

The unrenormalized ghost DSE reads \cite{Watson:2007vc}:
\begin{eqnarray}
\lefteqn{
\Gamma_{\overline{c}c}^{af}(k)=
\imath\delta^{af}\vec{k}^2
+\int\frac{d^4\w}{(2\pi)^4}\Gamma_{\overline{c}cAi}^{(0)abc}(k,\w-k,-\w)
}\nonumber\\&&\times
W_{\overline{c}c}^{bd}(k-\w)W_{AAij}^{ce}(\w)\Gamma_{\overline{c}cAj}^{dfe}(k-\w,-k,\w).
\end{eqnarray}
The ghost propagator ($W_{\overline{c}c}$) and the ghost proper two-point 
function ($\Gamma_{\overline{c}c}$) are decomposed as follows:
\begin{equation}
W_{\overline{c}c}^{ab}(k)=-\delta^{ab}\frac{\imath}{\vec{k}^2}D(\vec{k}^2),
\;\;\;\;
\Gamma_{\overline{c}c}^{ab}(k)=\delta^{ab}\imath\vec{k}^2\Gamma(\vec{k}^2).
\end{equation}
The dressing functions obey $D(\vec{k}^2)\Gamma(\vec{k}^2)=1$.  Notice that as a general result of the Slavnov-Taylor identities \cite{Watson:2008fb}, the ghost dressing functions are strictly independent of the energy.  With the vertex Ansatz, Eq.~(\ref{eq:vert}), one thus recognizes the temporal integral, Eq.~(\ref{eq:dat}), which can be replaced by the Gribov formula, Eq.~(\ref{eq:gribov}).  Introducing the ghost propagator renormalization coefficient $Z_c$ and writing in terms of the ghost dressing functions, the renormalized ghost DSE now reads
\begin{equation}
\Gamma\left(\vec{k}^2\right)=Z_c-\frac{g^2N_c}{16\pi^3}\!\!\int\!\!d\vec{\w}\frac{\sqrt{\vec{\w}^2}D\left((\vec{k}-\vec{\w})^2\right)}{(\vec{k}-\vec{\w})^2\sqrt{\vec{\w}^4+m^4}}\frac{k_ik_j}{\vec{k}^2}t_{ij}\left(\vec{\w}\right)
\end{equation}
where all quantities are implicitly renormalized at some scale $\mu$ and we have $N_c$ colors.

At one-loop perturbatively (obtained by setting $m=0$), using dimensional regularization and the renormalization scale $\mu$, one obtains the standard result \cite{Watson:2007vc}
\begin{equation}
\Gamma(x)=1+\frac{4}{3}\lambda N_c\ln{\left(x/\mu\right)}
\label{eq:pert0}
\end{equation}
where $x=\vec{k}^2$ and $\lambda=g^2/(4\pi)^2$.  One can see that for small $x$, $\Gamma(x)$ could become negative.  Gribov's original derivation of the formula, Eq.~(\ref{eq:gribov}), was based on expressly prohibiting such a situation \cite{Gribov:1977wm}.  We shall return to this topic later.  Naively resumming the result, Eq.~(\ref{eq:pert0}), gives the ultraviolet [UV] asymptotic behavior for the ghost:
\begin{equation}
D(x)=\left(\frac{x}{\mu}\right)^{\gamma_g},\;\;\gamma_g=-\frac{4}{3}\lambda N_c,
\label{eq:pert1}
\end{equation}
where $\gamma_g$ is (naively) the leading order expression for the ghost anomalous dimension.

One can also investigate the infrared [IR] behavior of the ghost DSE.  Using a UV-cutoff regularization and subtracting at $\mu$, the DSE has the form ($I$ is a function arising from the angular integration, see Ref.~\cite{Watson:2010cn} for the explicit form)
\begin{equation}
\Gamma(x)=\Gamma(\mu)-\lambda N_c\int_0^\Lambda\frac{dy}{y\Gamma(y)}\left[I(x,y;m)-I(\mu,y;m)\right].
\label{eq:dse0}
\end{equation}
The low $x$ and large $\Lambda$ scales separate because $\tilde{Z}_A=1$.  Assuming a powerlaw behavior $\Gamma(x)\sim(x/\mu)^\alpha$ as $x\rightarrow0$, one obtains the relationship
\begin{equation}
\left(\frac{x}{\mu}\right)^\alpha\sim\left(\frac{x}{\mu}\right)^{1-\alpha}\frac{\mu}{m^2}.
\end{equation}
This type of relationship is quite familiar in Landau gauge studies (e.g., \cite{Watson:2001yv}).  Counting powers of $x$, one sees that if there is an IR powerlaw, it will have the exponent $\alpha=1/2$.  However, counting powers of $\mu$ instead, one would have $\alpha=0$ which indicates an IR finite dressing function (and is intimately connected to the presence of the additional scale $m$ within the function $I$).

\begin{figure}[!t]
\resizebox{.8\columnwidth}{!}{\includegraphics{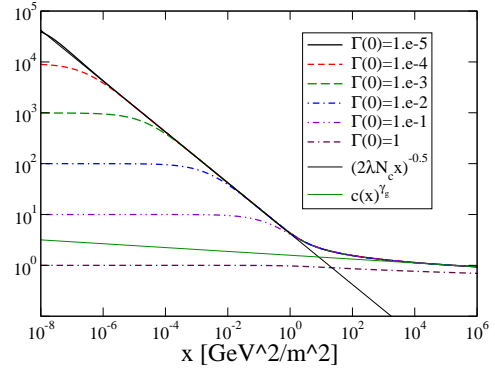}}
\caption{Results for the ghost propagator dressing function $D(x)$.  See text for details.}
\label{fig:ghost0}
\end{figure}

Equation (\ref{eq:dse0}) is not particularly suitable for numerical analysis because (as noted perturbatively) the subtraction at $\mu$ might lead to a negative result and numerical instabilities.  Taking inspiration from Gribov \cite{Gribov:1977wm} and demanding that $\Gamma(x)$ be positive for all $x>0$, one can subtract the equation at $x=0$.  Further scaling all dimensionful quantities with $m$ (e.g., $x\rightarrow xm^2$), the ghost DSE can then be written
\begin{equation}
\Gamma(x)=\Gamma(0)+\lambda N_c\int_0^\Lambda\frac{dy}{y\Gamma(y)}\left[\frac{4}{3}\frac{y}{\sqrt{1+y^2}}-I(x,y;1)\right].
\end{equation}
To evaluate this equation, there are two input parameters: $\lambda N_c$ ($\lambda=\alpha_s/(4\pi)$ with $\alpha_s=0.1187$ and $N_c=3$) and $\Gamma(0)\geq0$.  It is clear that $\Gamma(0)$ is not given by the DSE, but is a boundary condition for the solution.  Results for $D(x)=\Gamma(x)^{-1}$ are shown in Fig.~\ref{fig:ghost0} for various values of $\Gamma(0)$.  In the UV, the solutions with $\Gamma(0)\lesssim0.3$ lie on top of each other and agree asymptotically with the naively resummed perturbative result, Eq.~(\ref{eq:pert1}).  Going from high to low momenta, there is a characteristic behavior where the solutions for successively lower values of $\Gamma(0)$ follow the same curve further into the IR until `freezing' to the value $\Gamma(0)$ required by the boundary condition.  As $\Gamma(0)\rightarrow0$, one sees the emergence of the critical solution with the powerlaw exponent $\alpha=1/2$ as predicted.  Lattice results for the ghost propagator seem to rule out $\Gamma(0)\gtrsim0.2$ \cite{Watson:2010cn}.

\begin{figure}[!t]
\resizebox{.7\columnwidth}{!}{\includegraphics{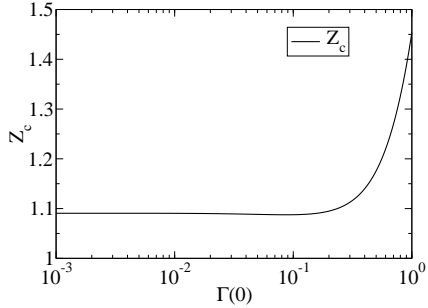}}
\caption{Results for the ghost renormalization coefficient as a function of $\Gamma(0)$.  See text for details.}
\label{fig:ghostz0}
\end{figure}

It is interesting to also calculate the renormalization coefficient for the ghost, $Z_c$.  In terms of the scaled variables, this is given by
\begin{equation}
Z_c=\Gamma(0)+\frac{4}{3}\lambda N_c\int_0^\Lambda\frac{dy}{\Gamma(y)\sqrt{1+y^2}}.
\end{equation}
Numerically, one can verify that $Z_c$ is independent of the momentum $x$, confirming the multiplicative renormalizability of the ghost DSE here.  Seeing the explicit dependence on $\Gamma(0)$ in the expression above, one can plot $Z_c$ as a function of $\Gamma(0)$ (with fixed $\Lambda$).  This is shown in Fig.~\ref{fig:ghostz0}.  Surprisingly, for $\Gamma(0)\lesssim0.3$, $Z_c$ shows almost no variation and indicating that the boundary condition is not connected to the renormalization.

After eliminating the possibility that $\Gamma(0)$ is connected to the renormalization, the characteristic pattern of the results exhibited in Fig.~\ref{fig:ghost0} can be naturally interpreted in terms of Gribov's gauge fixing ambiguity \cite{Gribov:1977wm} (indeed, this ambiguity was the origin of the condition $\Gamma(0)\geq0$).  Central to the gauge fixing is the Faddeev-Popov [FP] operator,
\begin{equation}
-\vec{\nabla}\!\cdot\!\vec{D}^{ab},\;\;\vec{D}^{ab}=\delta^{ab}\vec{\nabla}-gf^{acb}\vec{A}^c(x),
\end{equation}
and the Gribov problem arises when zero modes of the FP operator occur within the functional integral.  (Resolving the spatially independent zero modes that are particular to Coulomb gauge leads to a conserved and vanishing total color charge \cite{Reinhardt:2008pr}.)  Perturbatively, when one expands around $\vec{A}=0$, there are no zero modes but nonperturbatively in the IR, $\vec{A}$ can become large and zero modes appear.  With zero modes present, the gauge is not completely fixed and the functional integral contains an integration over the gauge group which when taking the expectation value of a gauge dependent quantity would lead to a suppression.  In this case, the suppression would presumably be a freezing at nonperturbative scales.  The DSE (at least in standard form) on the other hand does not change its form in the presence of the Gribov ambiguity - it merely relates the Green's functions to one another with their exact definition left implicit.  Thus, the common perturbative curve and the freezing of the solutions in the IR for finite $\Gamma(0)$ would appear to signal the presence of the Gribov ambiguity within the implicit definition of the functional integral for the ghost propagator.  Conversely, the $\Gamma(0)=0$ (critical) solution would appear to correspond to a more proper definition of the ghost.  Such a situation can be explicitly shown in $1+1$-dimensional Coulomb gauge \cite{Reinhardt:2008ij}.

The critical ghost solution can be tentatively linked to quark confinement \cite{Watson:2010cn}.  The Slavnov-Taylor identities \cite{Watson:2008fb} tell us that the dressing for the temporal gluon propagator has the form
\begin{equation}
D_{00}(k_0^2,\vec{k}^2)=\left[D(\vec{k}^2)\right]^2/\overline{\Gamma}_{AA}(k_0^2,\vec{k}^2)
\end{equation}
where $\overline{\Gamma}_{AA}$ is the longitudinal part of the spatial gluon polarization.  The renormalized DSE for $\overline{\Gamma}_{AA}$ would schematically be
\begin{equation}
\overline{\Gamma}_{AA}=Z_A+\sum_{i}Z_i\Sigma_i
\end{equation}
where each self-energy loop $\Sigma_i$ is multiplied by a vertex renormalization coefficient.  After subtracting at the scale $\mu$, this would be similar to the form of the ghost DSE, Eq.~(\ref{eq:dse0}), but with nontrivial $Z_i$ factors - only the ghost-gluon vertex (for which $\tilde{Z}_A=1$) has a trivial coefficient but since the corresponding loop involves energy independent ghosts, this loop must cancel (the infamous energy divergence problem of Coulomb gauge).  Thus, the separation of the external momentum and UV-cutoff scales necessary for the IR powerlaw analysis of Eq.~(\ref{eq:dse0}) is no longer applicable and one is left with the conclusion that $\overline{\Gamma}_{AA}$ must surely be a constant in the IR.  See Ref.~\cite{Watson:2010cn} for other arguments.  So with a critical ghost solution, $D_{00}\sim1/\vec{k}^2$ as $\vec{k}^2\rightarrow0$.  This is exactly the behavior one would expect in order to have a linear rising potential between quarks in Coulomb gauge \cite{Popovici:2010mb}.

\begin{theacknowledgments}
It is a pleasure to thank the organizers for an enjoyable conference.  Work supported by the DFG, contracts no. DFG-Re856/6-2,3.
\end{theacknowledgments}
\bibliography{proc}{}
\end{document}